\documentclass{article}

\usepackage{arxiv}

\usepackage[utf8]{inputenc} 
\usepackage[T1]{fontenc}    
\usepackage{hyperref}       
\usepackage{url}            
\usepackage{booktabs}       
\usepackage{amsfonts}       
\usepackage{nicefrac}       
\usepackage{microtype}      
\usepackage{lipsum}		
\usepackage{graphicx}
\usepackage{natbib}
\usepackage{doi}

\title{On Every Note a Griff:\\ 
Looking for a Useful Representation\\of Basso Continuo Performance Style}

\author{ \href{https://orcid.org/0009-0001-3201-9983}{\includegraphics[scale=0.06]{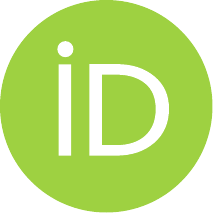}\hspace{1mm}Adam~Štefunko} \\
	Faculty of Mathematics and Physics\\
	Charles University\\
	Prague, Czech Republic \\
	\texttt{stefunko@ufal.mff.cuni.cz} \\
	\And
	\href{https://orcid.org/0000-0001-5770-7005}{\includegraphics[scale=0.06]{orcid.pdf}\hspace{1mm}Carlos~Eduardo~Cancino-Chacón} \\
	Institute of Computational Perception\\
	Johannes Kepler University\\
	Linz, Austria \\
	\texttt{carlos.cancino\_chacon@jku.at} \\
    \And
	\href{https://orcid.org/0000-0002-9207-567X}{\includegraphics[scale=0.06]{orcid.pdf}\hspace{1mm}Jan~Hajič~jr.} \\
	Faculty of Mathematics and Physics\\
	Charles University\\
	Prague, Czech Republic \\
	\texttt{hajicj@ufal.mff.cuni.cz} \\
}

\renewcommand{\shorttitle}{On Every Note a Griff}

\hypersetup{
pdftitle={On Every Note a Griff: 
Looking for a Useful Representation of Basso Continuo Performance Style},
pdfsubject={cs.SD},
pdfauthor={Adam Štefunko, Carlos Eduardo Cancino-Chacón, Jan Hajič jr.},
pdfkeywords={Basso continuo, Improvisation, Performance analysis},
}

\begin{document}
\maketitle

\begin{abstract}
    \textit{Basso continuo} is a baroque improvisatory accompaniment style which involves improvising multiple parts above a given bass line in a musical score on a harpsichord or organ. \textit{Basso continuo} is not merely a matter of history; moreover, it is a historically inspired living practice, and The Aligned Continuo Dataset (ACoRD) records the first sample of modern-day \textit{basso continuo} playing in the symbolic domain. This dataset, containing 175 MIDI recordings of 5 \textit{basso continuo} scores performed by 7 players, allows us to start observing and analyzing the variety that \textit{basso continuo} improvisation brings. A recently proposed \textit{basso continuo} performance-to-score alignment system provides a way of mapping improvised performance notes to score notes. In order to study aligned \textit{basso continuo} performances, we need an appropriate feature representation. We propose \textit{griff}, a representation inspired by historical \textit{basso continuo} treatises, that enables us to encode both pitch content and structure of a \textit{basso continuo} realization in a way invariant to transposition. \textit{Griffs} are directly extracted from aligned \textit{basso continuo} performances by grouping together performance notes aligned to the same score note in an onset order dependent way, and they provide meaningful tokens that form a feature space in which we can analyze \textit{basso continuo} performance styles. We statistically describe \textit{griffs} extracted from the ACoRD dataset recordings, and show in two experiments how \textit{griffs} can be used for statistical analysis of individuality of different players’ \textit{basso continuo} performance styles. We finally present an argument why it is desirable to preserve the structure of a \textit{basso continuo} improvisation in order to conduct a refined analysis of personal performance styles of individual \textit{basso continuo} practitioners, and why \textit{griffs} can provide a meaningful “historically informed” feature space worthy of a more robust empirical validation.
\end{abstract}


\section{Introduction}
\begin{figure}
	\centering
    \includegraphics[width=0.6\linewidth]{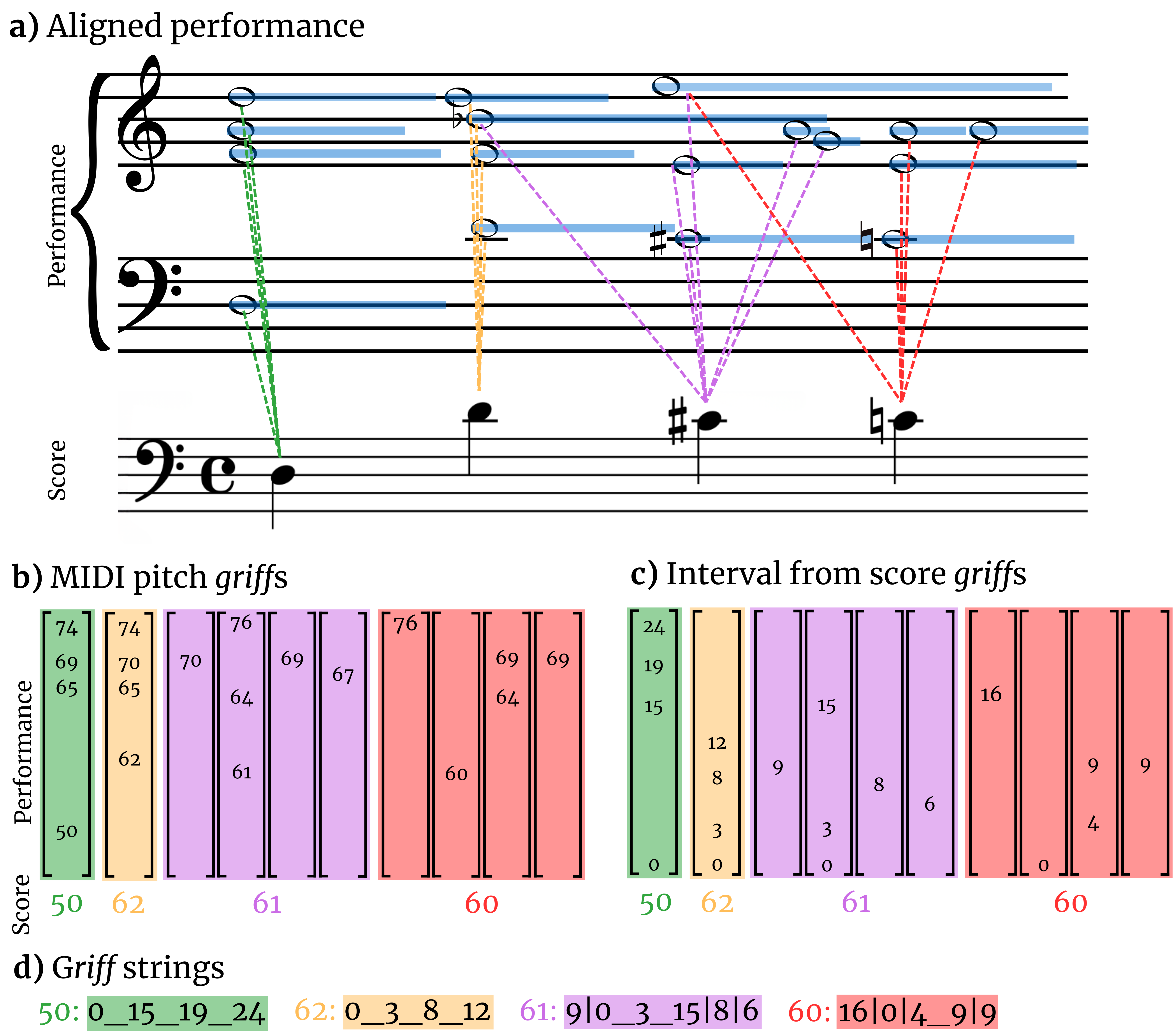}
	\caption{\textit{Griffs} are built using performances aligned to scores, and each score note forms its own \textit{griff} from performance notes aligned to it (\textbf{a}). MIDI pitches of performance notes are grouped to vectors based on the order and proximity of their onset times (\textbf{b}). A \textit{griff} is a sequence of these vectors. \textit{Griffs} converted to chromatic intervals from the score note pitch (\textbf{c}) are encoded as strings (\textbf{d}).}
	\label{fig:griff-extraction}
\end{figure}

\textit{Basso continuo}, the baroque improvisatory harmonic accompaniment style, requires the player to create a multiple-part improvisation based on a bass line from a musical score. This process is called \textit{basso continuo} realization. \textit{Continuo} practice has been resurrected in the 20\textsuperscript{th} century by the Historically-Informed Performance movement \citep{Christensen2002Continuo}, and a lot of harpsichord and organ players now play \textit{basso continuo} realizations regularly. However, despite its position at the attractive intersection between a well-understood musical style and orally transmitted improvisatory practice, little work has analyzed contemporary \textit{continuo} realizations.
The first sample of \textit{basso continuo} practice has been recorded in The Aligned Continuo Realization Dataset (ACoRD)\footnote{\href{http://hdl.handle.net/11234/1-5963}{\texttt{http://hdl.handle.net/11234/1-5963}}}, consisting of 175 MIDI recordings, totaling 6 hours of runtime spread across 66,967 notes. Each of 7 players realized 5 \textit{basso continuo} scores for 5 times. In order to enable a further analysis of these recordings, a performance-to-score alignment system was developed, leveraging the state-of-the-art symbolic alignment methods combined with preprocessing techniques and a time-concurrency dependent greedy algorithm \citep{Stefunko2025ACoRD}.
\textit{Basso continuo} alignment, albeit imperfect in some edge cases \citep{Chiruthapudi2025ContinuoAlignment}, establishes resources to start analyzing \textit{continuo} realizations. Now only features are missing. 
We propose a feature representation called \textit{griff}: configuration of tones played for a given score note (see Figure~\ref{fig:griff-extraction}). The word \textit{griff} stems from the \textit{basso continuo} treatise written by \citet{Muffat1699Regulae}, but the idea of an idiomatic chord note arrangement similar to guitar chord grips, but with many more options, is also present in other treatises \citep{Dandrieu1719Principes, Geminiani1756ThoroughBass}.

\textit{Griffs} become space dimensions for analyzing \textit{continuo} realization styles. At the same time, \textit{griffs} are a practically lossless encoding of pitch content of a realization. We hypothesize that, while the individual MIDI pitch and interval content of performances might be too coarse to study the individuality of the performance styles of different players, \textit{griffs} are features with appropriate resolution, as well as an encoding congruent with historical ways of thinking about \textit{continuo}.

\section{Building the \textit{Griffs}}

\textit{Griff} extraction begins with \textit{basso continuo} alignment mapping each score note to performance notes played to it. We propose two \textit{griff} representations:

\begin{itemize}
    \item \textit{Ordered}: a matrix of performance notes aligned to a score note in which the vertical axis represents pitches encoded either as MIDI numbers or as intervals from a given baseline, and the horizontal axis portraying the order of note onsets. It is very often desirable to join notes whose offsets differ only slightly in the same vector using a customizable quantization threshold parameter. All distinct pitches whose onsets appear within a window set by this parameter form a vector, pitches played after this window form the next vector, and a \textit{griff} is a sequence of such vectors. To distinguish between block chords and purposeful arpeggiation, we use 35\textit{ms} windows, a value based on previous research \citep{Flossmann2009PerformanceContext, Nakamura2014OPHMM}.
    \item \textit{Pooled}: All distinct pitches of performance notes aligned with the same score note form one vector. 
\end{itemize}

In order to work with an encoding invariant to musical transposition, we convert MIDI pitches to chromatic intervals from the score note. \textit{Griffs} are encoded as strings by separating intervals belonging to the same vector by an underscore, and neighboring vectors by a vertical line (see Figure~\ref{fig:griff-extraction}).

\textit{Griffs} should enable narrowing down the possible harmonic space of basso continuo realizations and bring out structured tonal content. \textit{Griffs} discard the absolute temporal relations between notes. However, this is beneficial because rhythmic differences expressed by absolute time tend to be variations of structurally equivalent patterns, so exact timing would obscure the equivalence of these patterns. Similar representations were used by \citet{Laaksonen2025Polyphony} for music pattern matching and by \citet{Peter2023SymbolicAlignment} in a \textit{DTW}-based alignment system.

\section{Analysis of the ACoRD Dataset Using \textit{Griffs}}
The following statistics and experiments are based on usage counts of different \textit{griff} types. In fact, most used \textit{griff} types are performance bass notes without harmony (724 times, 3.9\%) and score notes not aligned to any performance note (534 times, 2.9\%). However, since we are interested in analyzing the harmonic content of the dataset, we filter these two \textit{griff} types out.

\subsection{\textit{Griff} Statistics across the Dataset}
\begin{figure}
	\centering
    \includegraphics[width=0.6\linewidth]{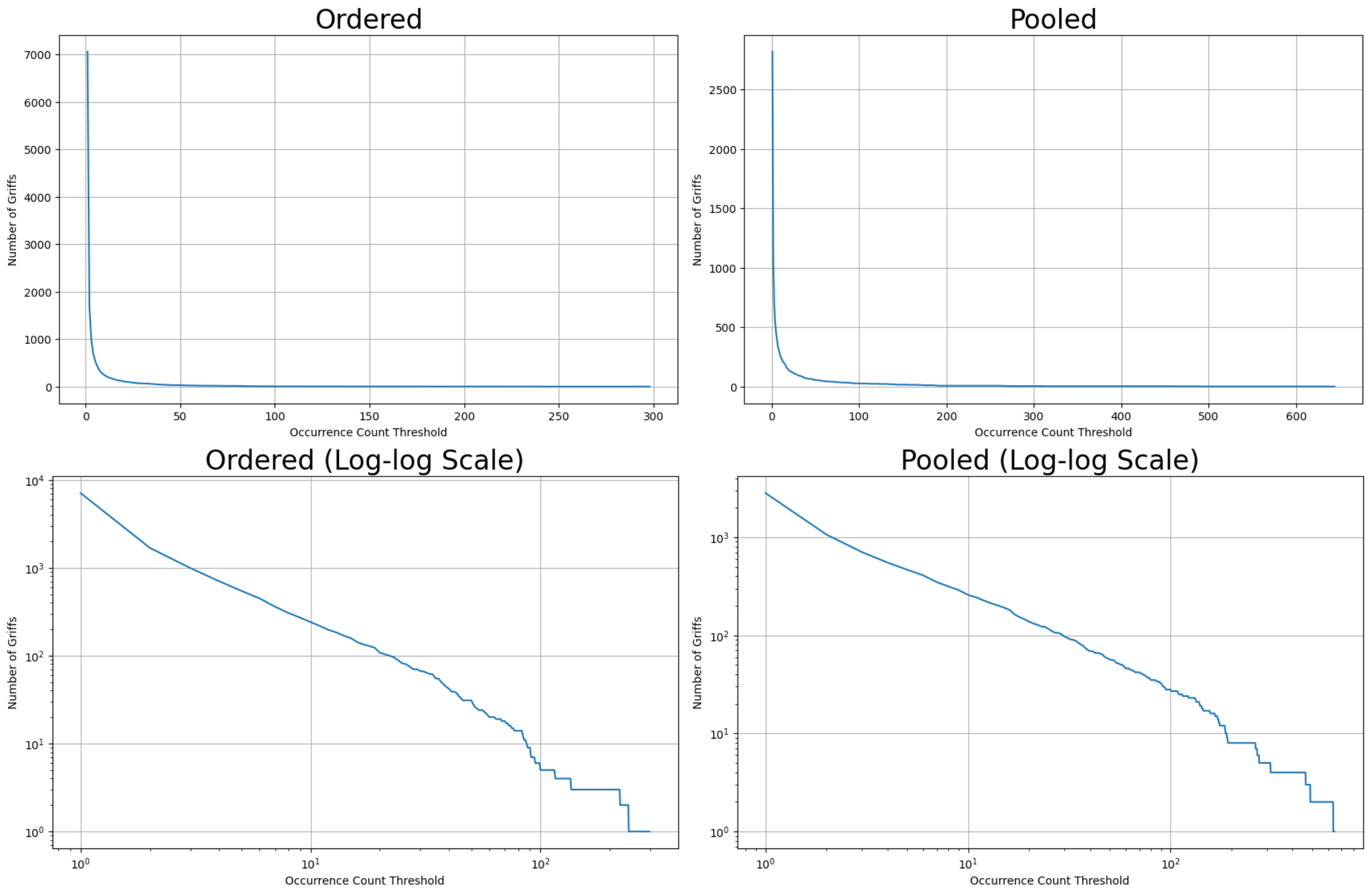}
	\caption{Distribution of different \textit{griff} types in both representations.}
	\label{fig:graphs}
\end{figure}

\begin{figure}
	\centering
    \includegraphics[width=0.6\linewidth]{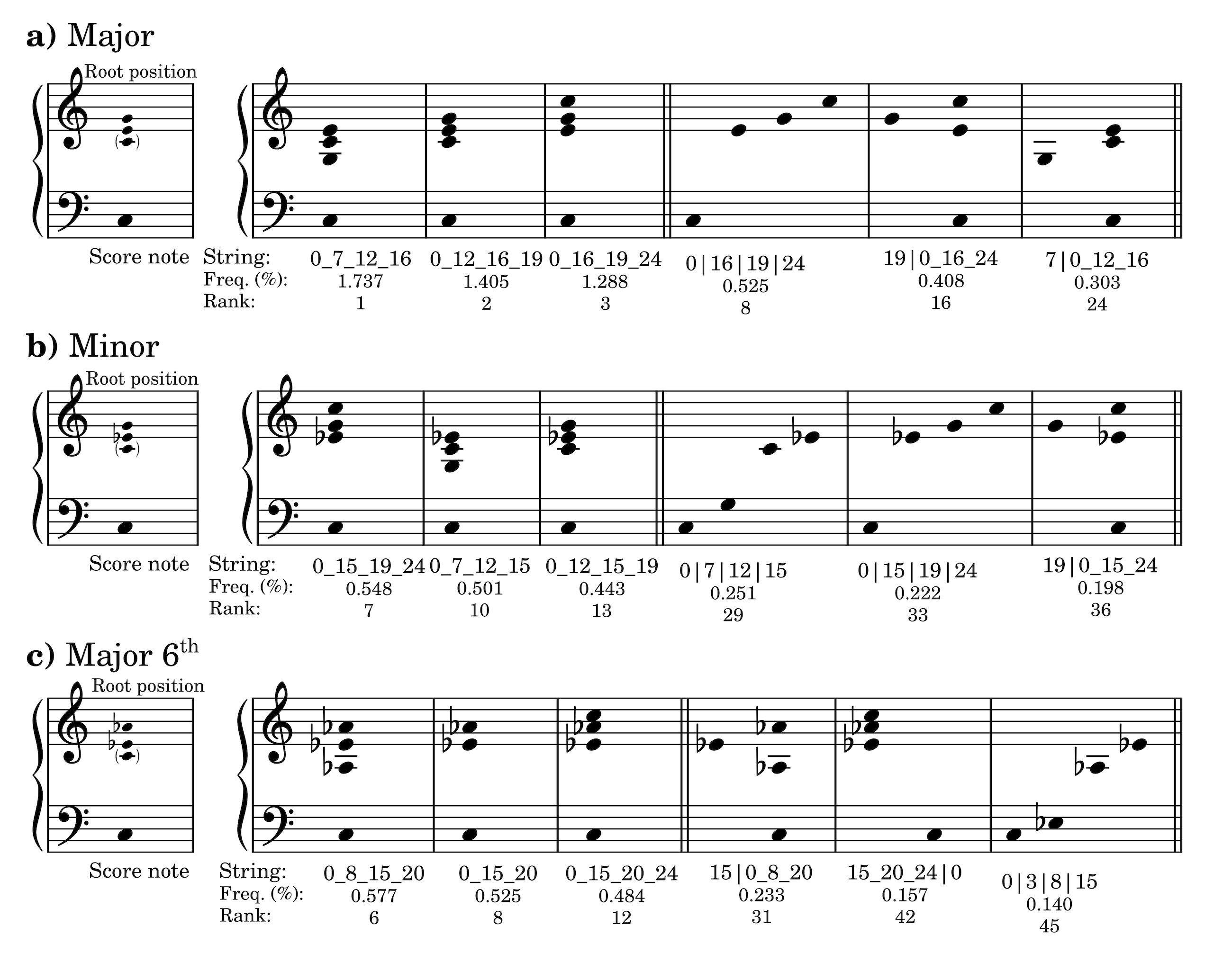}
	\caption{Three most occurring ordered \textit{griff} types without and with vertical lines belonging to the three most frequent chord types. The referential score note and a root position of the chords are shown. The figure contains string encoding of each \textit{griff}, frequency of occurrence and respective rank in the whole dataset.}
	\label{fig:most-used}
\end{figure}

Since ground-truth alignment data are available only for a limited number of performances, we use the best performing alignment method \textit{DualDTWNoteMatcher} \citep{Peter2023SymbolicAlignment} with trimming preprocessing \citep{Stefunko2025ACoRD}. We obtain 17,152 individual \textit{griffs} from the dataset after filtering. The basic statistics about the \textit{griffs} is shown in Table~\ref{tab:griff-statistics}. Distributions of both \textit{griff} representations are illustrated in Figure~\ref{fig:graphs}, and Figure~\ref{fig:most-used} shows examples of different chord patterns present in \textit{basso continuo} playing encoded as \textit{griffs}.

\begin{table}[h!]
\caption{Basic statistics of both \textit{griff} representations. \textit{Griff} occurrence count averaged over 17,152 \textit{griffs}.}
\centering
\begin{tabular}{@{}lll@{}}
\toprule
\textit{Griff} Representation           & Ordered & Pooled \\ \midrule
Total \textit{Griff} Types              & 7041    & 2817   \\
Average \textit{Griff} Occurrence Count & 2.44    & 6.09   \\ \bottomrule
\end{tabular}
\label{tab:griff-statistics}
\end{table}

\subsection{Experiments with Performance Profiles}
We investigate how individuality of players is reflected in the distribution of different \textit{griff} types across performances. We use performance \textit{griff} profiles, showing how often was each \textit{griff} type played in a performance. These profiles can be further aggregated across multiple performances. We do two experiments with \textit{griff} profiles:

\begin{itemize}
    \item A cumulative coverage indicating what percentage of \textit{griffs} in player’s performances is covered by \textit{k} \textit{griff} types they use most frequently.
    \item A measurement of how similar are player’s realizations of a score in comparison to other player’s realizations of the score in the space of \textit{griff} profiles. We measure cross-entropy\footnote{Any other similarity metric could be used, but since \textit{griff} profiles normalized over score length are categorical distributions, this is one natural choice.} of normalized \textit{griff} profiles on pairs of all 5 performances of a score by one player. We similarly measure cross-entropy on pairs of performances of two different players. Cross-entropies are computed for both \textit{griff} representations, and for frequencies of each chromatic interval to see how adding the structure to the intervallic content can help in measuring differences in performance styles. 
\end{itemize}

\subsection{Experiment results}
\begin{figure}
	\centering
    \includegraphics[width=0.6\linewidth]{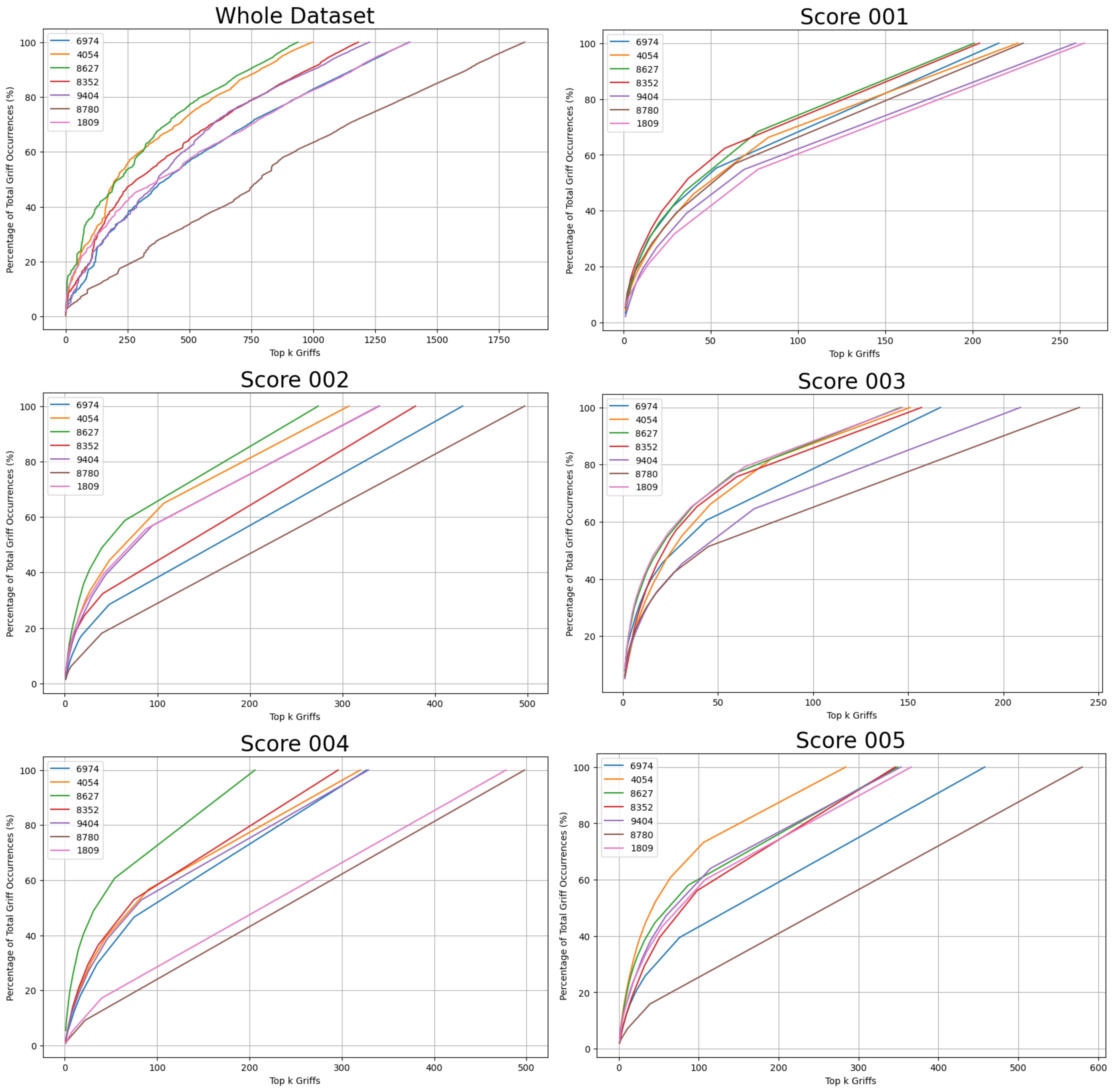}
	\caption{Cumulative coverage of ordered \textit{griff} types ranked from the most occurring to the least occurring ones for each player. The rightmost segment of a player’s line corresponds to \textit{griffs} used only once.}
	\label{fig:cumulative-coverage}
\end{figure}

Cumulative coverage indicates how inconsistent each player is in \textit{griff} usage, with results for the whole dataset and individual scores shown in Figure~\ref{fig:cumulative-coverage}. Cumulative coverage, however, does not reveal much information about individual styles because it does not take into account which \textit{griff} types each player uses most frequently.

\begin{figure}
	\centering
    \includegraphics[width=0.6\linewidth]{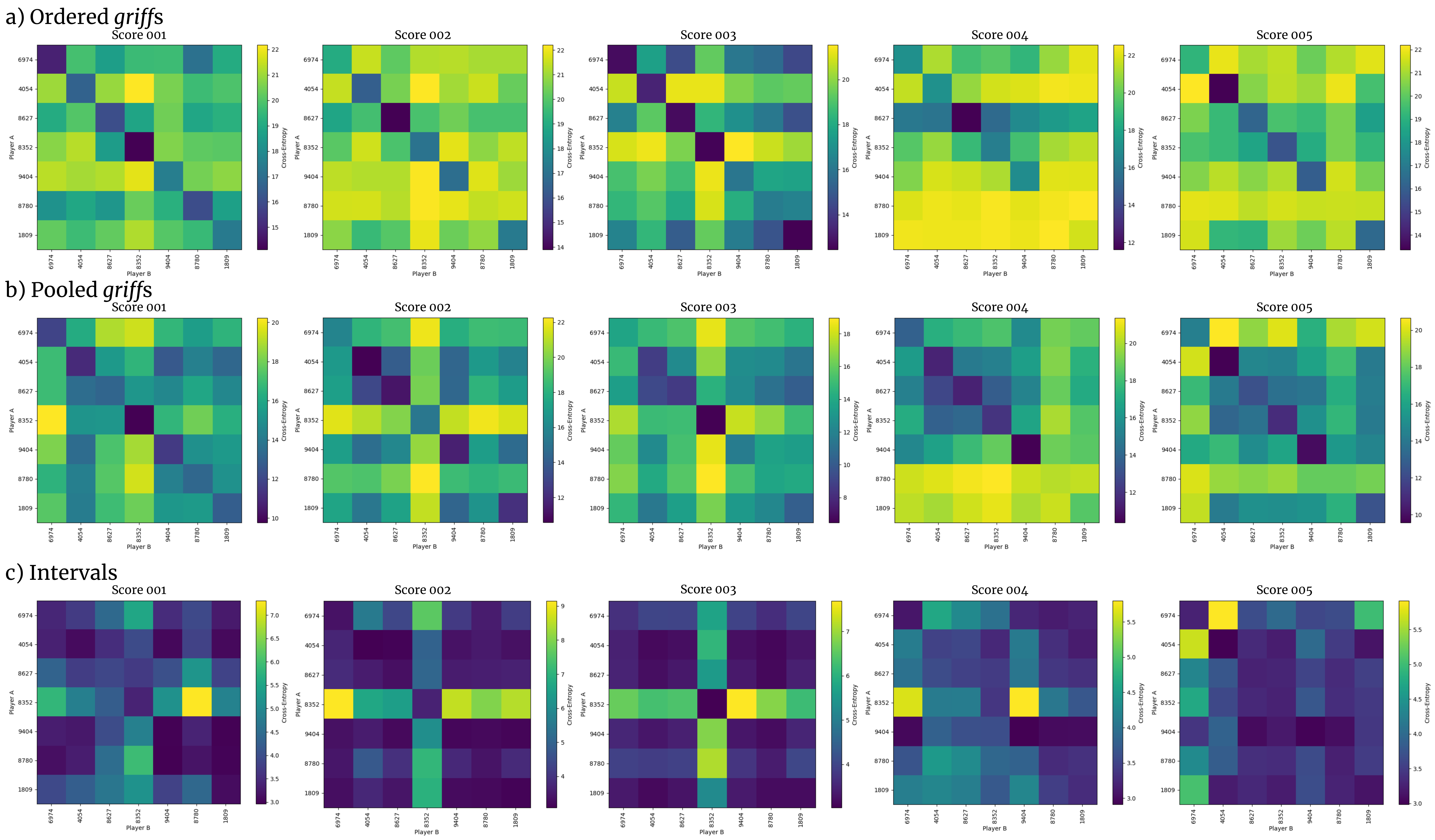}
	\caption{Similarity matrices of different players’ mean cross-entropies in three different representations for each score.}
	\label{fig:similarity-matrix}
\end{figure}

In order to study individuality of players, we need to find a level of representation that enables us to see variability across different players’ playing styles. Figure~\ref{fig:similarity-matrix} presents similarity matrices of mean pairwise cross-entropy values between performances of players for both \textit{griff} profile representations and, as a control, frequency profiles for intervals of individual realization notes from the score note (being a much less detailed representation).

The interval representation does not allow distinguishing between players. This is expected: harmonic constraints of \textit{continuo} realization heavily restrict the space of possible interval profiles for each score, so this representation is too coarse. In the \textit{griff} profile spaces, however, performances of the same player are more similar to each other than to performances of other players. If the \textit{griff} space was too fine-grained to be analytically useful, we would expect each performance pair to be approximately the same distance from each other. This is far from a complete analysis, and should not be taken as confirmation that there are individual \textit{continuo} realization styles yet, but we do take it as an indication that this “historically informed feature space” of \textit{griffs} is worth defining and worthy of a more robust empirical validation (such as with clustering and classification experiments, which are beyond the current scope).

\section{Conclusion}
In order to analyze and compare aligned \textit{basso continuo} recordings, it is necessary to come up with features that appropriately represent \textit{basso continuo} performances. \textit{Griffs} seem to be a historically-informed and empirically promising way of representing aligned basso continuo performances with little loss, and thus provide meaningful tokens for analysis of continuo as a living improvisatory tradition.

\bibliographystyle{unsrtnat}
\bibliography{references}  






\end{document}